\documentclass[twocolumn]{revtex4-2}
\usepackage{epsfig}
\usepackage{graphics}
\usepackage{amssymb}
\usepackage{textcomp}
\usepackage{amsmath}
\usepackage[dvipsnames]{xcolor}
\usepackage{subfigure}

\def\eq#1{(\ref{#1})}

\begin{document}

\title{On the initial state of the universe and the time arrow}

\author{Carlos Silva} 
\email{carlosalex.phys@gmail.com}
\affiliation{Instituto Federal de Educa\c{c}\~{a}o Ci\^{e}ncia e Tecnologia do Cear\'{a} (IFCE),\\ Campus Tiangu\'{a} - Av. Tabeli\~{a}o Luiz Nogueira de Lima, s/n - Santo Ant\^{o}nio, Tiangu\'{a} - CE, Brasil}

\author{Renan Aragão}
\email{renangomezzz@gmail.com }
\affiliation{Departamento de Física, Universidade Federal de Campina Grande, Caixa Postal 10071, Campina Grande, 58429-900, Paraíba, Brasil}

\author{Francisco A. Brito}
\email{fabrito@df.ufcg.edu.br}
\affiliation{Departamento de Física, Universidade Federal de Campina Grande, Caixa Postal 10071, Campina Grande, 58429-900, Paraíba, Brasil}


\begin{abstract}

In this paper, we investigate an important issue addressed by several approaches to quantum cosmology, like that based on loop quantum gravity and string theory: the existence of an arrow of time driving the cosmological evolution according to the generalized second law of thermodynamics. In this way,  by considering quantum corrections to the Friedmann equations,  we find out that
for such a cosmological arrow of time to exist, our universe must have occupied a state of negative entropy during its very early evolution. This opens a new, and interesting, perspective on the beginning of the cosmos: that our universe has emerged not from a Big Bang, but from a Dirac sea of negative entropy.


\end{abstract}


\maketitle

\section{Introduction}

Despite the advances in the description of the cosmos promoted by General Relativity, several issues about our universe remain obscure.
Among such issues, is the existence of an arrow of time driving the universe's evolution from a low entropy state to a high entropy one. It corresponds to our universe's obedience to the generalized second law of thermodynamics (GSL).

The existence of an arrow of time in the cosmos is related to another crucial question about its evolution: what is the initial state of our universe? Possibly, such issues must be fixed by a theory of quantum cosmology based on a forthcoming theory of quantum gravity. However, amidst the efforts in this
direction, to obtain a quantum description of the cosmos has been a tough challenge.

Despite this, attempts have been made to shed light on the issue of the initial state of the universe in the context of quantum gravity, by proposing some possible scenarios for its beginning. For example, according to loop quantum gravity (LQG), the initial state of the cosmos must not correspond to a singularity, as proposed by general relativity,  but to a quantum bounce where the universe can transit from an earlier contraction phase to the current expansion evolution \cite{Bojowald:2001xe, Bojowald:2006da, Taveras:2008ke}.
Other approaches based on string theory - the other leading proposal to quantum gravity -  introduce similar scenarios \cite{Khoury:2001wf, Steinhardt:2002ih, Veneziano:1997kx, Buchbinder:2007at}.


The issue of the validity of the GSL in our universe has been widely discussed in the literature, as that related to string cosmology \cite{Brustein:1999ay, Brustein:1999ua}, loop quantum cosmology (LQC) \cite{MohseniSadjadi:2010nu, Sadjadi:2012wg}, and other approaches \cite{Wall:2010jtc, Radicella:2010ss,  Bousso:2015eda, Ben-Dayan:2020pbg, Odintsov:2024hzu}. However, the nature of the mechanism that leads the cosmos to obey such a law remains obscure. Particularly, it has been demonstrated,  in the context of LQG, that it is not possible to say that our universe obeys the GSL if one applies the usual entropy-area formula provided by this approach to study the early cosmological evolution \cite{Sadjadi:2012wg}


On the other hand, recently, it has been demonstrated that quantum-corrected Friedmann equations can be obtained from thermodynamical arguments \cite{Silva:2015qna}, by using a quantum-corrected entropy-area formula introduced in the context of loop quantum black holes \cite{Modesto:2008im, Modesto:2009ve, s.hossenfelder-grqc12020412, Carr:2011pr}. This approach proved capable of reproducing similar equations to those proposed by LQC.

In the present article, we shall address the issue of the validity of the GSL in the context of the scenario proposed in \cite{Silva:2015qna}, by considering the case of a flat universe like ours. In this case, we shall utilize holographic
approaches and insights from LQG to address the validity of the GSL during the early stages of cosmic evolution.

As we shall see, one can get obedience to the GSL in the scenario introduced in \cite{Silva:2015qna}. However, it becomes necessary to completely break with the ideas proposed till now to the beginning of the cosmos by admitting that the universe has evolved from an initial negative entropy state to a positive one during its very early evolution. It opens the possibility that our universe has emerged not from a Big Bang, nor it comes from a quantum bounce, but from a kind of negative entropy Dirac sea \cite{Song:2014}.

The paper is organized as follows: in section  \eq{sec2}, we shall review the results introduced in \cite{Silva:2015qna}. In section \eq{sec3}, we shall adress the universe's early evalution in face of the results introduced in \cite{Silva:2015qna}. In section \eq{sec4}, we shall address the validity of the GSL in the context of such results. Section \eq{conc} is devoted to discussions and conclusions.
Throughtout the paper we have used $c = G = \hslash = k_{B} = 1$, unless otherwise stated.

\section{Quantum corrected Friedmann equations from a thermodynamical argument}  \label{sec2}

Despite its complexity, according to the cosmological principle, our universe
can be considered, at a very large scale, as
homogeneous and isotropic.
Based on this simplifying assumption, the Friedmann
equations correspond to the ones that govern the expansion of the universe in the framework of the scientific paradigm proposed by General Relativity.

Friedmann equations were first derived in $1922$ by introducing the  Friedmann-Lema\^{\i}tre-Robertson-Walker (FLRW) metric in Einstein's field equations of gravitation,  considering a perfect fluid as the source \cite{friedmann-zphys10}.

The Friedmann equations of a uniform cosmology are typically written in the form

\begin{equation}
\dot{H} -  \frac{k}{a^{2}} =  -4\pi(\rho_{tot} - p_{tot}),
\end{equation}

\noindent and 
\begin{equation}
H^{2} + \frac{k}{a^{2}} = \frac{8\pi}{3}\rho\; . \label{classical-friedmann}
\end{equation}

In the equations above, $H$ is the Hubble parameter, and $a(t)$ is the dimensionless scale factor of the universe. Moreover, $\rho$ and $p$ are respectively the energy density and pressure of the perfect fluid. The dimensionless constant $k$ corresponds to the spatial curvature of the universe. The Hubble parameter is defined as $H = \dot{a}/a$, and $a_{0}$ is the scale factor
of the universe at some canonical time $t_{0}$.


On the other hand, Friedmann equations must incorporate quantum corrections to explain the evolution
of the universe in the stages close to the Big Bang singularity, where spacetime must have a quantum
behavior. Moreover, to describe the evolution of the universe, cosmological equations like the Friedmann ones must include elements to describe the thermodynamical aspect of the cosmos.

%
%
%
%
%
%
%
%

In this sense, it has been demonstrated that quantum-corrected Friedmann equations can be derived by the use of thermodynamical arguments \cite{Silva:2015qna},  by assuming that the entropy associated with the universe's apparent horizon
is related to its area by the modified entropy-area relation introduced in the context of loop quantum black holes \cite{Modesto:2008im, Modesto:2009ve, s.hossenfelder-grqc12020412, Carr:2011pr}.  In the present section, we shall review the main steps to find out such results.

In this way, we have that a FLRW universe is described by the following metric

\begin{eqnarray}
ds^{2} &=& - dt^{2} + a(t)^{2}\Big(\frac{dr^{2}}{1- kr^{2}} + r^{2}d\Omega^{2}_{2}\Big) \nonumber \\
        &=& h_{ab}dx^{a}dx^{b} + \tilde{r}^{2}d\Omega^{2}_{2}\label{frw-metric} \; ,
\end{eqnarray}

\noindent where $h_{ab} = diag (-1, a^{2}/(1- kr^{2}))$ and $\tilde{r} = a(t)r$.


Now, let us suppose that the energy-momentum tensor $T_{\mu\nu}$ of the matter in the universe has the form of a perfect
fluid:

\begin{equation}
T_{\mu\nu} = (\rho + p)U_{\mu}U_{\nu} + p g_{\mu\nu}\;.
\end{equation}

The energy conservation law leads to the continuity equation

\begin{equation}
\dot{\rho} + 3H(\rho + p) = 0\;. \label{continuity-eq}
\end{equation}

To analyze the thermodynamical evolution of the universe, we shall consider its apparent horizon, whose radius is given by

\begin{equation}
\tilde{r}_{A} = \frac{1}{\sqrt{H^{2} + k/a^{2}}} \label{ap.hor-rad}\;. 
\end{equation}

In this case, we shall consider a heat flux crossing the universe's apparent horizon, which can be calculated by the use of  the first law of thermodynamics  \cite{Cai:2005ra, Cai:2006rs}:

\begin{equation}
dE = A \psi_{a} dx^{a} + WdV\; ,
\end{equation} 

\noindent where $E$ is the Misner-Sharp energy, $A$ is the apparent horizon area, $V$ is the volume enclosed by the universe's apparent horizon. Moreover, 
the work density $W$ and the energy-supply vector $\psi$ are defined as

\begin{equation}
W = -\frac{1}{2}T^{ab}h_{ab}\; ;
\end{equation}
\noindent and

\begin{equation}
\psi_{a} = T^{b}_{a}\partial_{b}\tilde{r} + W\partial_{a}\tilde{r} \; .
\end{equation}

\noindent In the present scenario, we shall have

\begin{equation}
W = \frac{1}{2}(\rho - p)\;,
\end{equation}

\noindent and
\begin{equation}
\psi_{a} =  \left(- \frac{1}{2}(\rho + p)H\tilde{r}, \frac{1}{2}(\rho + p)a \right) \; .
\end{equation}

From the results above, we can compute the amount of energy going through 
the apparent horizon during the time interval
$dt$ as \cite{Cai:2005ra}

\begin{equation}
dQ = - A\psi = A(\rho+p)H\tilde{r}_{A}dt \; , \label{heat-var}
\end{equation}

\noindent which we shall consider as corresponding to a heat flux.

Moreover, as it has been emphasized by Cai et al \cite{Cai:2008ys}, the horizon temperature is
completely determined by the spacetime metric,
independently of gravity theories.
In this way, the temperature associated with the apparent horizon is given by

\begin{equation}
T = \frac{1}{2\pi\tilde{r}_{A}}, \label{ap-hor-temp}
\end{equation}

\noindent which was obtained in the reference \cite{Cai:2008gw} through tunneling methods.

On the other hand, the horizon entropy depends on the gravity theory considered. In this way, the choice of the entropy-area relation attributed to the universe's apparent horizon will constitute a central point of our analysis.

In this case, it has been demonstrated in \cite{Silva:2015qna} that non-singular Friedmann equations similar to the LQC ones can be obtained from a quantum corrected Bekenstein-Hawking formula introduced at first in the context of self-dual black holes  \cite{Modesto:2008im, Modesto:2009ve, s.hossenfelder-grqc12020412, Carr:2011pr}. 

Such an entropy-area formula is given by


\begin{equation}
S = \pm \frac{\sqrt{A^{2} - A_{min}^{2}}}{4\sigma}, \label{entropy-area}
\end{equation}

\noindent where $A_{min}$ corresponds to a minimal area parameter.
Moreover, $\sigma = 1 - P^{2}$, where $P$ is the so-called polymeric function which appears in the LQG quantization techniques.

An interesting fact about the entropy \eq{entropy-area} is that it can be negative. In the context of self-dual black holes, it occurs when the black hole enters a subPlanckian phase which is dominated by higher curvature effects which lead to a repulsive gravity regime where the black hole singularity is replaced by a bounce \cite{Modesto:2009ve}. In fact, it has been pointed out in other contexts that negative entropy may be attributed to higher curvature effects  \cite{Cvetic:2001bk}, as those leading to repulsive gravity regimes \cite{Kehagias:2015ata}. We note that repulsive gravity is intimately related to polymer quantization effects coming from LQG \cite{Singh:2015jus}.

Based on these considerations about the thermodynamics of the universe's apparent horizon, now we shall use the Clausius relation
$dQ = TdS$. In this case, we shall apply the results in the Eqs. \eq{heat-var} and \eq{ap-hor-temp}, together with the following result for the apparent horizon entropy variation, obtained from Eq. \eq{entropy-area}:

\begin{equation}
dS = \pm \frac{A}{4\sigma\sqrt{A^{2} - A_{min}^{2}}} dA.
\end{equation}

We obtain

\begin{equation}
(\rho + p)H\tilde{r}_{A} = \pm \frac{1}{2\pi\tilde{r}_{A}}\frac{1}{4\sigma\sqrt{A^{2} - A^{2}_{min}}}\frac{dA}{dt}, \label{clausius1}
\end{equation}

\noindent where

\begin{equation}
\frac{dA}{dt} = - \frac{A^{2}}{2\pi}H\left(\dot{H} - \frac{k}{a^{2}}\right).
\end{equation}

\noindent In order to get the equation above, we have used also the relation

\begin{equation}
\dot{\tilde{r}}_{A} = -H\tilde{r}^{3}_{A}\Big(\dot{H} - \frac{k}{a^{2}}\Big)\;.
\end{equation}

By gluing it together the  Eq. \eq{clausius1}, we obtain

\begin{equation}
\dot{H} - \frac{k}{a^{2}} =  \mp 4\pi \sigma \frac{\sqrt{A^{2}-A_{min}^{2}}}{A}(\rho + p)\;. 
\label{friedmann1}
\end{equation}

\noindent Now, by applying the continuity Eq. \eq{continuity-eq}, we can find

\begin{eqnarray}
\frac{8\pi}{3}d\rho &=& \pm \frac{1}{\sigma}\frac{A}{\sqrt{A^{2} - A_{min}^{2}}} d(H^{2}+k/a^{2}) \nonumber \\
                                           &=&  \mp \frac{1}{\sigma}\frac{4\pi}{A\sqrt{A^{2} - A_{min}^{2}}}dA \; 
\end{eqnarray}

\vspace{5mm}

\noindent where we have used the Eq. \eq{ap.hor-rad}.

\vspace{5mm}

Integrating the equation above yields

\begin{equation}
\Theta = \pm\Big[\frac{2A_{min}}{3}\sigma\rho - \alpha\Big] = \arccos(A_{min}/A) \;, \label{theta-eq}
\end{equation}

\noindent where we must have $- \pi/2 \leq \Theta \leq \pi/2$, since $A_{min}/A \geq 1$. In the Eq. above, $\alpha$ is a constant phase.

\vspace{5mm}

One can obtain the following Friedmann equation from the Eq. \eq{theta-eq}:

\begin{equation}
H^2 + \frac{k}{a^2} = \frac{4\pi}{A_{min}}\cos(\Theta)\;. \label{friedmann2}
\end{equation}

The Eqs. \eq{friedmann1} and \eq{friedmann2} consist of the quantum version of the Friedmann equations obtained from the entropy-area relation \eq{entropy-area}. As we can see, the quantum corrections in these equations imply in a quantum effective density term
which is a harmonic function of the classical density. Such a result has also been found out in the context of the AdS/CFT correspondence \cite{Silva:2020bnn}, where it has opened the doors for an interesting discussion on the nature of spacetime in quantum gravity \cite{Silva:2023ieb}.

It is interesting to note that, by considering the case of a flat universe like ours,  from Eq. \eq{friedmann2} and the continuity equation \eq{continuity-eq}, we can obtain the Raychaudhuri equation:

\begin{equation}
\dot{H} = 4\pi (p + \rho)\sin\left(\frac{2A_{min}}{3}\sigma\rho - \alpha \right)\;. \label{q-ray-eq}
\end{equation}

\noindent Such a result is independent of the sign of the entropy \eq{entropy-area}.

\section{Universe's early evolution} \label{sec3}

The scenario introduced in \cite{Silva:2015qna} allows the discussion of three regimes to the very early evolution of a flat universe, by considering the validity of the null energy condition, in a way that $\rho+p\geq 0$:

\vspace{5mm}

\noindent (i) Superinflation where $\dot{H} > 0$. Such a regime occurs when

\begin{equation}
\rho > \frac{3\alpha}{2A_{min}\sigma}.
\end{equation}

\vspace{5mm}

\noindent (ii) Ordinary inflation where $\dot{H} < 0$, which occurs when

\begin{equation}
\rho < \frac{3\alpha}{2A_{min}\sigma}.
\end{equation}

\vspace{5mm}

\noindent (iii) Transition time where $\dot{H} = 0$, which occurs when

\begin{equation}
\rho = \frac{3\alpha}{2A_{min}\sigma}.
\end{equation}

\vspace{5mm}

In this way, a superinflationary regime arises in the scenario introduced in \cite{Silva:2015qna}, at the early cosmological evolution, as a consequence of higher curvature effects present in the Raychaudhuri equation \eq{q-ray-eq} when $\rho > \frac{3\alpha}{2A_{min}}$. However, as the universe expands, and its density becomes lower, the higher curvature effects become less relevant so that the cosmos leaves the superinflationary phase, passes through the transition time, and enters the ordinary inflation phase when $\rho < \frac{3\alpha}{2A_{min}}$.


\subsection{Early entropic evolution of the cosmos.} \label{inflaton-solution}

By considering the results above, it is possible to trace out the evolution of the universe's entropy by taking into account the following results coming from thermodynamics of the apparent horizon \cite{Cai:2005ra,Cai:2006rs}
\begin{equation}
dU=-TdS\to d(\rho V)=\rho dV + Vd\rho  = - TdS \; , \label{first-law}
\end{equation}
where $V=4\pi \tilde{r}_A^3/3\sim H^{-3}(t)$ is the volume related to the apparent horizon. 

Now, by considering that, near the transition point between superinflationary and inflationary regimes, $H(t) \approx H(t_{0}) \approx \text{const.}$, where $t_{0}$ corresponds to the transition time, we can take $dV \sim \dot{H}\sim 0$ and rewrite Eq. \eq{first-law} as
\begin{equation}
TdS = - Vd\rho,
\end{equation}
which gives us
\begin{equation}
S = -\int\frac{Vd\rho}{T}= -\frac{4\pi}{3}\int\frac{\tilde{r}_A^3d\rho}{T}.
\end{equation}

In this way, by taking $T=1/2\pi\tilde{r}_A$ (Eq.~\eq{ap-hor-temp}) and $\tilde r_A=H(t)^{-1}$ (Eq.~\eq{ap.hor-rad} for $k=0$) we obtain
\begin{equation}
S(t)= -\frac{8\pi^2}{3}\int H^{-4}(t)\frac{d\rho}{dt}dt,
\end{equation}
the universe's entropy, as a function of time.

\subsection{Early inflationary evolution and its impact to cosmological entropy.} \label{inflaton-solution}

To better understand the significance of the results in the previous section, we shall solve the cosmological equation found in the case where the early evolution of the universe is dominated by a scalar field (inflation).

In such a scenario, the matter density and pressure can be written as

\begin{eqnarray}
p = \frac{1}{2}\dot{\phi}^{ 2} + V(\phi); \nonumber \\
\rho = \frac{1}{2}\dot{\phi}^{ 2} - V(\phi)\;.
\end{eqnarray}

In this case, the continuity equation \eq{continuity-eq} will be written as

\begin{eqnarray}
\ddot{\phi} + 3H\dot{\phi} + \frac{\partial V}{\partial\phi} = 0.
\end{eqnarray}

By considering the slow-roll approximation $\ddot{\phi} \ll 3H\dot{H}$ ($\rho \approx V(\phi)$ and $\frac{1}{2}\dot{\phi}^{2} \ll V(\phi)$), we find:

\begin{equation}
3H\dot{\phi} + \frac{\partial V}{\partial\phi} = 0; \label{inflaton-cont-eq}
\end{equation}

\noindent and 

\begin{equation}
H^{2} = \frac{4\pi}{A_{min}}\cos\left[\alpha\left(\frac{V(\phi)}{\rho_{0}} - 1 \right)\right]. \label{inflaton-fried-eq}
\end{equation}

By substituting \eq{inflaton-fried-eq} in \eq{inflaton-cont-eq}, we get

\begin{equation}
\dot{\phi} = - \frac{1}{3}\frac{1}{\sqrt{\frac{4\pi}{A_{min}}}\cos^{1/2}\left[\alpha\left(\frac{V(\phi)}{\rho_{0}} - 1 \right)\right]}\frac{\partial V(\phi)}{\partial \phi} \;,
\end{equation}

\noindent which is, in general, difficult to integrate.

However, let us choose the flat (steep) potential \cite{Cook:2015vqa}

\begin{equation}
V(\phi) = \lambda \phi.
\end{equation}

In this case,

\begin{equation}
\dot{\phi} = - \frac{1}{3}\frac{\lambda}{\sqrt{\frac{4\pi}{A_{min}}}\cos^{1/2}\left[\alpha\left(\frac{\lambda\phi}{\rho_{0}} - 1 \right)\right]}\frac{\partial V(\phi)}{\partial \phi} \;,
\end{equation}

By integrating, we obtain

\begin{equation}
\int d\phi \sqrt{\cos{\left[\alpha\left(\frac{\lambda\phi}{\rho_{0}} - 1 \right)\right]}} = -\frac{\lambda}{3}\sqrt{\frac{A_{min}}{4\pi}}t + cte \;,
\end{equation}

\noindent which gives us

\begin{equation}
-\frac{2}{a}E\left[\frac{1}{2}(b - a\phi) \mid 2\right] = -\frac{\lambda}{3}\sqrt{\frac{A_{min}}{4\pi}}t + cte \;,
\end{equation}

\noindent where $E(x \mid 2)$ corresponds to the Eliptic function, and

\begin{equation}
a = \frac{\alpha\lambda}{\rho_{0}} \;,\; b = \alpha.
\end{equation}

For $\phi$ or $a$ very small, we have the first two terms

\begin{equation}
-\frac{2}{a}E\left(\frac{b}{2}\right) + \sqrt{\cos{\alpha}}\phi + ... = -\frac{\lambda}{3}\sqrt{\frac{A_{min}}{4\pi}}t + cte,
\end{equation}

\noindent which can be simplified, by redefining some constants, in a way that:

\begin{equation}
\sqrt{\cos{\alpha}}\phi = -\frac{\lambda}{3}\sqrt{\frac{A_{min}}{4\pi}}t + \phi_{0}\;,
\end{equation}

or

\begin{equation}
\phi(t) = \phi_{0} -\frac{\lambda}{3\sqrt{\cos{\alpha}}}\sqrt{\frac{A_{min}}{4\pi}}t \;.
\end{equation}

In this case, the Hubble parameter turns to depend on time as

\begin{eqnarray*}
H(t) = \sqrt{\frac{4\pi}{A_{min}}}\cos^{1/2}\Big[\alpha\Big(\frac{\lambda}{\rho_{0}}\Big(\phi_{0} - \frac{\lambda}{3\sqrt{\cos{\alpha}}}\sqrt{\frac{A_{min}}{4\pi}}t\Big)&&  \\ - 1\Big)\Big]&&\;.
\end{eqnarray*}

Moreover, the energy density, $\rho(t) = \frac{1}{2}\dot{\phi}^{2} + V(\phi) \approx V(\phi) = \lambda\phi (t)$, will be given by:

\begin{equation}
\rho(t) =  \lambda \left(\phi_{0} -\frac{\lambda}{3\sqrt{\cos{\alpha}}}\sqrt{\frac{A_{min}}{4\pi}}t\right) \;.
\end{equation}

\begin{figure}[htb]
     \centering  
      \hspace{5mm}\includegraphics[width=8cm, height=6cm]{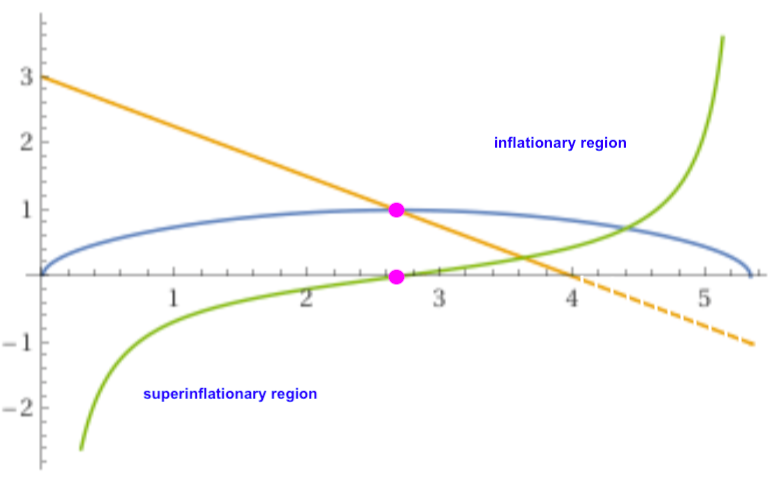}
      \caption{Time evolution of the universe density $\rho(t)$ (orange), the Hubble parameter $H(t)$ (blue), and the universe entropy $S(t)$ (green). A superinflationary regime arises during the very early evolution of the cosmos due to higher curvature effects present in the Raychaudhuri equation \eq{q-ray-eq}. During this regime, the universe's entropy assumes negative values. However, as the universe expands and its density becomes lower, the higher curvature effects become less relevant so that the cosmos leaves the superinflationary phase, passes through the transition time, and enters the ordinary inflationary regime, where the universe's entropy becomes positive. For $\alpha=\pi/4, \phi_0=2, \lambda=\rho_0=1.5, A_{\rm min}=4\pi$. } \label{fig1}
\end{figure}

 By considering such results, we can plot the dependence of the Hubble parameter, as well as of the matter density, and entropy with time as displayed in Fig.~\ref{fig1}. As we can see, we have the presence of a superinflationary era during the very early stages of the universe's evolution, where $\dot{H} > 0$, as predicted in the discussions above. During this regime, the universe's entropy is negative. As the universe's density becomes lower, the ordinary inflationary regime takes place, and the universe passes to a regime of positive entropy.

Some comments are in order related to the methods used in the present section. In this way, we have traced our equations by considering a regime of thermodynamic equilibrium. On the other hand, during the superinflationary period, the universe is far from the de Sitter attractor, being subject to global non-equilibrium. Despite this, local quasi-equilibrium of the region inside the universe's apparent horizon can be implemented, which leads to a negative entropy inside the apparent horizon. In this case, by considering the cosmological principle, other regions of the universe must equally have negative entropy during this regime, and it is possible to say that the universe, as a whole, has passed through a negative entropy state during superinflation.

\subsection{Relationship with usual semiclassical LQC.}

As it has been demonstrated in \cite{Silva:2015qna}, it is possible to contrast the results obtained above with usual semiclassical LQC. In this sense, let us expand the right-hand side of the Eq. \eq{friedmann2} as

\begin{eqnarray}
H^{2} + \frac{k}{a^{2}} &\sim&\frac{1}{\gamma \sqrt{3}}\cos (\alpha)  + \frac{8\pi}{3}\sigma \sin (\alpha) \rho \nonumber \\
&-&  \frac{32\pi^{2}}{9}\gamma\sigma^{2}\sqrt{3}\rho^{2}\cos (\alpha)  \nonumber \\
&+& \textrm{ higher derivative terms}\; , \label{eq-expanded}
\end{eqnarray}

\noindent where we have used $A_{min} = 4\pi \gamma \sqrt{3}$ \cite{Perez:2004hj}, with $\gamma$ the Barbero-Imirzi parameter.

It is possible to rewrite the  Eq. \eq{eq-expanded} as

\begin{eqnarray}
H^{2} &+& \frac{k}{a^{2}} =  \frac{8\pi}{3}\rho_{tot}\Big( 1 - \frac{\rho_{tot}}{\rho_{c}}\Big) \nonumber \\ 
          &+& \textrm{ higher derivative terms}, \label{lqc-equation}
\end{eqnarray}

\vspace{5mm}

\noindent where $\rho_{tot} = \rho + \frac{\Lambda}{8\pi}$, and $\Lambda$ is a cosmological constant.

In order to obtain the equation above, the following conditions must be satisfied:

\begin{equation}
\rho_{c} = \frac{\sqrt{3}}{4\pi\gamma\sigma^{2}\cos (\alpha)}\; ,
\end{equation}

\begin{equation}
1 - \frac{\tilde{\Lambda}}{\rho_{c}} = \sigma \sin (\alpha)\; ,
\end{equation}

\begin{equation}
\tilde{\Lambda}\Big( 1 - \frac{\tilde{\Lambda}}{\rho_{c}}\Big) = \frac{3 \cos (\alpha)}{8\pi\gamma\sqrt{3}} \; ,
\end{equation}

\noindent where $\tilde{\Lambda} = \frac{\Lambda}{8\pi G}$.

In the context of the results found out in \cite{Silva:2015qna},  one can also obtain the Raychaudhuri equation from the time derivative of the Eq. \eq{lqc-equation}, which by the use of the continuity equation \eq{continuity-eq}, gives us

\begin{eqnarray}
\dot{H} -  \frac{k}{a^{2}} &=&  -4\pi(\rho_{tot} - p_{tot})\Big( 1 - \frac{\rho_{tot}}{\rho_{c}}\Big) \nonumber \\ 
                                           &+& \textrm{ higher derivative terms}\;,
\end{eqnarray}

\noindent where $p_{tot} = p - \frac{\Lambda}{8\pi}$.

Moreover, the following result for the cosmological constant is obtained:

\begin{eqnarray}
\Lambda =  8\pi G\tilde{\Lambda}  &&= 4\pi\rho_{c}\Big[1 - \Big(1 - \frac{3}{4\pi^{2}\gamma^{2}\rho_{c}^{2}}\Big)^{\frac{1}{4}}\Big] \nonumber \\
                                                    &&\approx \frac{3}{4\pi\gamma^{2}\rho_{c}}. \label {cosm.constant}
\end{eqnarray}

\vspace{1cm}

From such results, we obtain for the phase constant $\alpha$, by considering $\rho_{c} \gg \tilde{\Lambda}$

\begin{equation}
\alpha \approx \arccos{\left(\frac{8\pi\gamma}{\sqrt{3}}\tilde{\Lambda}\right)}.
\end{equation}

In this way, the parameter $\alpha$ can be fixed by the value of the cosmological constant in our universe.

In this way, the value of the cosmological constant in the Eq. \eq{cosm.constant} can be used to fix both the value of the universe's critical density, and the phase constant $\alpha$. It is because, as it has been argued in \cite{Silva:2015qna}, the LQC point of view about the cosmological constant is that it can not be theoretically explained but must be considered a fundamental constant in nature like Newton's gravitational constant or Planck's constant \cite{Wilson-Ewing:2016yan}. In this case, by taking the observed value of the cosmological constant ($\sim 10^{-122}$),  it is possible to argue that the universe's critical density must be superPlanckian.

In this way,  the complete LQC semiclassical dynamics, plus a cosmological constant, and high derivative terms,  can be obtained from a thermodynamical argument based on the LQBH entropy-area relation \eq{entropy-area}.
It has traced a possible way to understand LQC equations from a holographic point of view.

We note that the higher derivative terms above, which correspond to higher curvature ones, have not been discussed suitably in the LQC models yet, since such models are not based on a derivation from a covariant quantum theory \cite{Bojowald:2019ckm}.
However, in the present analysis, such terms must be related to the presence of a superinflationary regime in the very early evolution of the cosmos. Moreover, higher curvature terms must be related to one of the most important results we shall obtain in the present paper: that the universe must have occupied a state of negative entropy during the early stages of its evolution.

%
%
%

%

\section{The generalized second law of thermodynamics during the early stages of cosmological evolution.}  \label{sec4}

By obtaining quantum-corrected Friedmann equations from a thermodynamics description, the important issue related to the existence of an arrow of time in our universe arises.

Particularly, the issue of the validity of a GSL in the context of LQC was analyzed by Sadjadi \cite{Sadjadi:2012wg}, where the usual LQG entropy-area relation was used. However, in such a context, it is found that the GSL was violated during the so-called superinflation era.

In this section, we shall investigate the issue of the GSL in the context of the results introduced in \cite{Silva:2015qna} by considering the case of a flat universe. Such a discussion will be dictated by the regimes for the very early evolution of the cosmos given by the Raychaudhuri equation \eq{q-ray-eq} introduced in Sec. \eq{sec3}: the super inflationary regime, the transition time, and the ordinary inflationary regime. Such regimes have also been analyzed in the context of usual LQC \cite{Sadjadi:2012wg}.


\vspace{2mm}

In this sense, we have that the variation of the universe's entropy with time can be written as

\begin{equation}
\dot{S_{t}} = \dot{S}_{h} + \dot{S}_{in} \geq 0,
\end{equation}

\noindent where $\dot{S}_{h}$ corresponds to the variation of the entropy of the universe's apparent horizon, and $\dot{S}_{in}$ corresponds to the variation of the matter plus radiation content inside it.


In this way,  by using the entropy-area formula \eq{entropy-area} for the universe's horizon, one obtains

\begin{equation}
\dot {S}_{h} = -  sgn (S_{h})\frac{1}{2} \frac{[A^{2}-A^{2}_{min}]^{\frac{1}{2}}}{4\sigma} 2A \dot{A}, \label{horizon-entropy-var}
\end{equation}

\noindent where $A = 4 \pi \tilde{r}^{2}_{A}$ is the area of the universe's apparent horizon, whose radius is given by

\begin{equation}
\tilde{r}_{A}= \frac{1}{H}, \label{ap.hor-rad2}
\end{equation}

\noindent  in the case of a flat universe.

By considering the Eqs \eq{horizon-entropy-var} and \eq{ap.hor-rad2}, we obtain


\begin{equation}
\dot{S}_{h}= - sgn (S_{h})\frac{2\pi}{\sigma} \left[ 1 - \frac{A^{2}_{min}H^{4}}{16\pi^{2}}\right]^{-\frac{1}{2}}\frac{\dot{H}}{H^{3}} .
\end{equation}

Moreover, the variation rate of the entropy associated with  the matter plus radiation content inside the universe's apparent horizon is given by

\begin{equation}
\dot{S} _{in}= -4\pi \frac{(p+\rho)}{T H^{2}} \left[  1+   \frac{ \dot{H} }{H^{2}}            \right].
\end{equation}

In this way, by adding the contributions found out above, the total variation rate for the universe's entropy will be given by

\begin{eqnarray}\label{Taxa de variação da entropia total}
\dot{S}_{tot}= &&- sgn (S_{h})\frac{2\pi}{\sigma}\left[ 1 - \frac{A^{2}_{min}H^{4}}{16\pi^{2}}\right]^{-\frac{1}{2}}\frac{\dot{H}}{H^{3}} \nonumber \\
 &&-4\pi \frac{(p+\rho)}{T H^{2}} \left[  1+   \frac{ \dot{H} }{H^{2}} \right].
\end{eqnarray}

Now, let us analyze the validity of the GSL in the present context by considering at first the superinflationary regime, where  $\dot{H}>0$. In this way, by considering that the universe horizon remains in a negative entropy state during this epoch, we shall have $sgn (S_{h}) <  0$. In this case, it is possible to obtain $\dot{S}_{tot} \geq 0$ during this regime if the contribution from the apparent horizon to the universe's entropy variation is larger than the contribution from matter.

We note that for the case where the universe is filled only by a scalar field, as we have addressed in Sec. \eq{inflaton-solution}, for what $p = -\rho$, the condition above is automatically satisfied. On the other hand, by considering a more general case where other kinds of fields are present, such a condition turns out to be equivalent to imposing a lower bound on the minimal area parameter:

\begin{equation}
A_{min} \geq \frac{4\pi}{H^{2}}\Big[1 - \frac{\dot{H}^{2}H^{4}}{4\pi^{2}\sigma^{2}(p+\rho)^{2}(H^{2}+\dot{H})^{2}}\Big]^{1/2}, \label{min-area-lbound}
\end{equation}

\noindent where we have used the Eqs. \eq{ap-hor-temp} and \eq{ap.hor-rad2} to obtain such a result.

Eq. \eq{min-area-lbound} also  imposes that

\begin{equation}
\sigma \geq \frac{\dot{H}H^{2}}{2\pi(p+\rho)(H^{2} + \dot{H})},
\end{equation}

\noindent and

\begin{equation}
P^{2}  \leq 1- \frac{\dot{H}H^{2}}{2\pi(p+\rho)(H^{2} + \dot{H})}.
\end{equation}

Now, by considering the transition time, we shall have

\begin{eqnarray*}
- sgn (S_{h}) \left[ 1 - \frac{A^{2}_{min}H^{4}}{16\pi^{2}}\right]^{-\frac{1}{2}}\frac{\dot{H}}{H^{3}}  \geq 2\sigma \frac{(p+\rho)}{T H^{2}} \left[  1+   \frac{ \dot{H} }{H^{2}} \right] ,
\end{eqnarray*}

\noindent or yet

\begin{eqnarray*}
\frac{\dot{H}}{H^{3}} \geq - sgn(S_{H})2\sigma \left[ 1 - \frac{A^{2}_{min}H^{4}}{16\pi^{2}}\right]^{\frac{1}{2}} \frac{(p+\rho)}{T H^{2}} \left[  1+   \frac{ \dot{H} }{H^{2}} \right].
\end{eqnarray*}

However, at the transition time, $\dot{H}=0$. In this case, we obtain

\begin{equation}
0 \geq - sgn(S_{H})2\sigma \left[ 1 - \frac{A^{2}_{min}H^{4}}{16\pi^{2}}\right]^{\frac{1}{2}}\frac{(p+\rho)}{T H^{2}}.
\end{equation}

Again, for the case where the universe is filled only by a scalar field as we have addressed in Sec. \eq{inflaton-solution}, the condition above is automatically satisfied. On the other hand, by considering a more general case where other kinds of fields are present, the GSL remains valid during the transition time if one has a positive or null entropy associated with the universe's apparent horizon during this epoch.

By considering the case where we have a null entropy, one obtains
 
\begin{equation}
\frac{\sigma}{2\pi} \left[ 1 - \frac{A^{2}_{min}H^{4}(t_{0})}{16\pi^{2}}\right]^{\frac{1}{2}} 4\pi \frac{(p+\rho)}{T H^{2}} = 0.
\end{equation}

As one can observe, the contribution to the entropy variation rate from matter plus radiation content cannot be equal to zero, in a way that the horizon entropy variation rate must vanish, i.e.,

\begin{equation}
\frac{2\pi}{\sigma} \left[ 1 - \frac{A^{2}_{min}H^{4}(t_{0})}{16\pi^{2}}\right]^{\frac{1}{2}} = 0,
\end{equation}

\noindent consequently,

\begin{equation}
A_{min} = \frac{4\pi}{H^{2}(t_{0})}.
\end{equation}

At the transition time, $H(t_{0}) = \sqrt {\frac{2\pi\rho_{c}}{3}}$. Thus,  we obtain the minimal area parameter as follows,

\begin{equation}
A_{min} = \frac{6}{\rho_{c}}.
\end{equation}

We note that the bound above can be taken as compatible with the LQG prediction for $A_{min}$, which in turn must depend on the value of the Barbero-Immirzi parameter.
However, we highlight that there is no consensus in the definition of $A_{min}$, nor $\gamma$, nor even $\rho_c$ in the LQG literature. See Refs.~\cite{Dittrich:2007th,Vyas:2022etz,Mercuri:2009vk,Bojowald:2008ik,Dzierzak:2008dy, Malkiewicz:2009zd}.

At last, during the ordinary inflationary regime, we have that $\dot{H}<0$. In this case, by considering that the contribution from the apparent horizon to the entropy variation rate is bigger than the contribution from matter, one can obtain a positive variation rate to the universe's entropy
from \eq{Taxa de variação da entropia total}, by simply considering that
$S_{h}>0$ during the inflationary epoch.

From the results above, we observe that to have the validity of the GSL during the regimes analyzed in the present paper, it is necessary that the universe has passed from a negative entropy state in the superinflationary regime to a positive entropy state in the regime of ordinary inflation. Moreover,  spacetime entropy must overweigh the matter plus radiation one, implying in this case in a lower bound to the minimal area parameter.

\section{Discussions and conclusions} \label{conc}

In this paper, we have investigated the validity of the GSL in the context of the approach introduced in \cite{Silva:2015qna}, where quantum corrected Friedmann equations have been obtained. Such equations reproduce the semiclassical LQC scenario plus higher curvature terms and a cosmological constant.

We have analyzed three regimes for the evolution of the universe as given by the Raychaudhuri equation \eq{q-ray-eq}: the superinflationary regime, the transition time, and the ordinary inflationary regime. 
Such regimes have been also analyzed in the context of LQC \cite{Sadjadi:2012wg}.



It is possible to argue that the GSL remains valid during the regimes above if one considers that there is a lower bound to the minimal area parameter, which is in agreement with one hopes from a quantum gravity scenario. Moreover, the universe must have passed from a negative entropy state during the superinflationary regime to a positive entropy state in the epoch of ordinary inflation.

The negative values for the universe's entropy during the superinflationary regime must be related to higher curvature effects which dominate this phase of the early cosmological evolution \cite{Cvetic:2001bk, Kehagias:2015ata}. However, as the universe expands and its density becomes lower, the higher curvature effects become less relevant, and the universe's entropy becomes positive.

The present results challenge the usual proposals by LQG and string theory to quantum cosmology, that the universe has passed from an earlier contraction regime to the current expansion one. It is because, in such a case, the universe should have passed from a positive entropy state before the bounce to a negative entropy one, after (during) this process, which would violate the GSL. In this way, a possible scenario that arises in the light of the present results is that the universe has emerged from some negative entropy state without a prior evolution.

Even though only a complete theory of quantum gravity can offer us a full explanation of this kind of phenomenon related to the universe's initial state, the results found out in the present paper provide a prospect on the beginning of the cosmos that brings an additional task to programs for the quantization of gravity: to explain how negative entropy states can appear in their frameworks.

In this sense, even though we have obtained Friedmann equations similar to that obtained in the context of
LQC, the additional higher curvature terms, which lead to negative entropy states at the beginning of the cosmos, are not well understood in the LQC context, since its models are not based on a derivation from a covariant quantum theory \cite{Bojowald:2019ckm}.

The issue of negative entropy remains an additional challenge when one considers the quantization of general relativity since it corresponds to a quantum concept that has no parallel in classical physics.
For example, we note that the concept of negative entropy does not match the standard definition of thermodynamic entropy $S = k_{B}\ln\Omega$, which is used
to measure the entropy of a system in terms of the number of its degrees of freedom.

In this way, we must highlight that the formula above gives the value of the absolute entropy of the system, which can not be negative. On the other hand, negative entropy values can be attributed to some physical system by considering not its absolute entropy, but its conditional entropy measured in relation to an ancilla \cite{Cerf:1995sa, Rio:2011}.

In this way, by considering the existence of a negative entropy state to the universe at its beginning, the first interpretation of such a result is that negative entropy is due to the entanglement of our universe with some ancilla system.

However, what kind of system could be used as an ancilla to the universe? Such a system would need to be separate from the universe itself. However, it does not make sense in a cosmological scenario, since it is nothing beyond the cosmos.

On the other hand, an interesting scenario has been introduced in \cite{Song:2014} which has brought an alternative interpretation to the meaning of negative entropy. In this case, it has been proposed that there must exist a kind of Dirac sea of negative entropy filling up the system, which plays the role of the ancilla. In such a scenario, a system with positive entropy can emerge from the Dirac sea of negative entropy, leaving behind a ‘hole’ that corresponds to the ancilla quantum state \cite{Song:2014}. It opens the possibility that our universe has emerged not from a singularity, nor does it come from a previous contracting phase, but from a Dirac sea of negative entropy.

However, the specific description of such a Dirac sea of negative entropy will stay on a full theory of quantum gravity, and as far as we know, there is no clear scenario in order to obtain such a theory.

In such a scenario, one could be tempted to ask: what could be a hole in such a Dirac sea of negative entropy? Could it be an excess of positive entropy due to quantum gravity corrections, as it occurs in the case of a logarithmic quantum corrections to entropy, as appears for example in \cite{Majumdar:2000pr}?


A possible guide to such discussions is the observation that the results found out in the present paper are deeply rooted in the quantum-corrected Raychaudhuri's equation \eq{q-ray-eq} found out in Sec. \eq{sec3}. Note that even the entropy-area formula \eq{entropy-area} can obtained from such an equation by inverting our arguments between Eqs. \eq{entropy-area} and  \eq{q-ray-eq}.

Raychaudhuri’s equation studies the evolution of the congruence of timelike or null geodesics by describing how the expansion of a bundle of geodesics behaves in spacetime and can be seen as the origin of the second law of thermodynamics in the context of black holes \cite{Bhattacharya:2020wdl}.

In this way, the central question in our analysis must be: how do quantum corrections to the Raychaudhuri equation affect the congruences of geodesics so that negative entropy states can be obtained and the GSL obeyed?

In this sense, by considering the behavior of spacetime at very small scales,  one can attempt to quantize aspects of spacetime geometry itself, such as the metric tensor, or geodesic congruences.  Recent attempts to achieve this quantization have been
grounded in the geometric quantization approach \cite{Gotay:1980xk, Isenberg:1981md, Woodhouse:1980de,  Woodhouse:1992de, Bates:1997kc, Nair:2016ufy, Carosso:2018ihc, Moshayedi:2020spz, Camosso:2020zmr,  Berman:2022acl, Wernli:2023pib, Tawfik:2024itt, NasserTawfik:2024afw, NasserTawfik:2024abc}. Such a formalism provides a direct bridge between classical and quantum physics, allowing for a formal and systematic way of transitioning from a classical phase space to a quantum Hilbert space.

By considering the dependence of the results introduced in the present work on the Raychaudhuri equation \eq{q-ray-eq}, geometric quantization could be useful to describe how quantum fluctuations of spacetime geometry affect the behavior of geodesics, which in turn impacts the evolution of geodesic congruences described by such an equation.
Interestingly,  geometric quantization has shown to be useful both in LQG \cite{Conrady:2009px} and string theory \cite{Yu:1989jaa, Yu:1989jb, Popov:1993qzx, Merkulov:1992xm} contexts.

Moreover, we note that in the context of the geometrical quantization approach, the choice of how classical phase space is structured may affect the way subsystems of the quantum system are correlated at the microscopic level. In this case, it is possible to inquire if some choice of the classical phase space configuration of general relativity could lead to entangled states characterized by negative entropy configurations. Further investigations must shed more light on the possibility of using geometric quantization to clarify the significance of the results in the present paper.

It is important to highlight that, in the present work, we have introduced a new scenario to the initial conditions of the universe, which must result in observational imprints, for example, in the universe perturbation spectra and in the CMB.

Moreover, since the physics developed in our manuscript is related to the properties of self-dual black holes, some observational signatures of such physics may be present in our universe.

In fact, the results found out in the present paper enlarge the doors for the concept of negative entropy in cosmology. In this case, by considering structure formation in a universe where negative entropy plays a role, possibly black holes described by the self-dual solution can play a role. Such black holes can leave imprints in the CMB (through the production of gravitational waves, or Hawking radiation), or yet in the large-scale structure of the cosmos, by considering the influence of black holes in this process.
For example, self-dual black holes have been pointed out as candidates do dark matter \cite{Modesto:2009ve}. Further analysis must give us the way observational imprints will appear in a scenario where the universe began in a negative entropy state.


Such a possibility for the beginning of the cosmos proposed in the present paper must be investigated in other non-singular cosmological scenarios, even in those we do not need a phase of standard inflation during the early stages of the universe \cite{Creminelli:2006xe, Creminelli:2007aq, Creminelli:2010ba, Cai:2012va, Cai:2014jla, Brandenberger:2016vhg}.

\section*{Acknowledgements}
\noindent \parbox{9cm}{The authors acknowledge the anonymous referees for their useful comments and criticism. Francisco \;A.\;Brito acknowledges support from CNPq (Grant number 309092/2022-1) and also CNPq/PRONEX/FAPESQ-PB (Grant number 165/2018), for partial financial support.}

\end{document}